\documentclass[a4,amsmath,amssymb]{revtex4}
\usepackage{amsmath}
\usepackage{amssymb}
\usepackage{graphicx}% Include figure files
\usepackage{bm}% bold math
\def\plotX{8.3cm}
\def\elec{e}
\def\eqnn#1{(\ref{eq:#1})}
\def\sect#1{Sect.~\ref{sec:#1}}

\def\figno#1{Fig.~\ref{fig:#1}}

\def\Im{{\rm \,Im}}
\def\ns{{\cal N}}

\def\vev#1{\langle #1\rangle}
\def\Im{{\rm \,Im}}

\def\kb{k_{\scriptscriptstyle B}}
\def\omr{\omega_{R}}
\def\kr{k_{R}}

\def\kMax{{k_{\rm max}}}

\def\bsize{b}
\def\bpower{{\cal P}}
\def\beff{\kappa}
\def\NA{N\!A}
\begin{document}
\title{Spectral properties of thermal fluctuations on simple liquid surfaces
  below shot noise levels
}
\author{ Kenichiro Aoki\footnote{E--mail:~{\tt ken@phys-h.keio.ac.jp}.
    Supported in part by the Grant--in--Aid for Scientific Research
    (\#20540279) from the Ministry of Education, Culture, Sports,
    Science and Technology of Japan.}  and Takahisa
  Mitsui\footnote{E--mail:~{\tt mitsui@hc.cc.keio.ac.jp}.}  }
\affiliation{Dept. of Physics, Hiyoshi, Keio University, Yokohama
  223--8521, Japan}
\begin{abstract}
    We study the spectral properties of thermal fluctuations on simple
    liquid surfaces, sometimes called ripplons.  Analytical properties
    of the spectral function are investigated and are shown to be
    composed of regions with simple analytic behavior with respect to
    the frequency or the wave number.  The derived expressions are
    compared to spectral measurements performed orders of magnitude
    below shot noise levels, which is achieved using a novel noise
    reduction method.  The agreement between the theory of thermal
    surface fluctuations and the experiment is found to be excellent.
\end{abstract}
\vspace{3mm}
\maketitle
\section{Introduction}
\label{sec:intro}
Thermal fluctuations are ubiquitous and exist for practically
everything we see and touch. However, they tend to be too small to be
observed or measured directly, except under special circumstances.
Phenomena in which thermal fluctuations can be examined directly in
non-exotic materials are surface fluctuations of liquids, sometimes
called ``ripplons''\cite{ripplon}.  Using surface light scattering,
thermal surface fluctuations of liquids have been studied for some
time\cite{ripplonExp}. Other direct thermal fluctuation measurements
include high power interferometry of mirror surfaces\cite{mirror} and
fluctuations of surfaces with exceptionally low surface
tension\cite{giant}.

In this work, we detect reflected light from surfaces to measure their
inclinations\cite{opticalLever,Tay,am1} and apply this to thermal
surface fluctuations of simple liquids. The measurements are somewhat
complementary to the spectral measurements performed at specific
wavelengths. The measurements require less power, can be performed on
smaller samples and can be applied to strongly viscous
fluids. Furthermore, the measurement system can be simpler.  While
shot noise is often regarded as an unavoidable limitation in such
measurements, we show that such is not the case.  The measurements
involve novel methods that allow us to measure the spectrum directly,
down to several orders of magnitude below the shot noise level. As we
explain, this principle is not limited to thermal noise measurements
nor to surface light scattering measurements. The experimental results
agree with the theory quite well. In the process, we elucidate the
simple analytic behavior of the spectra and show how it is reflected
in the full spectrum, which can be seen in experiments.

The dispersion relation for surface waves on a simple liquid surface
can derived from the Navier-Stokes equation, when the
viscosity of the liquid can be ignored, as
\begin{equation}
    \label{eq:dispersion}
    \omega(k)=\sqrt{{\sigma k^3\over \rho}+gk} \qquad.
\end{equation}
Here, $g$ is the gravitational acceleration, $\rho,\sigma$ are the
density and the surface tension of the liquid, $k$ is the wave number
and $\omega$ is the (angular) frequency. The viscosity becomes more
important for shorter wavelengths and dissipation will play an
essential role below. Gravitational effects are more important for
longer wavelengths and is negligible for wavelengths much smaller than
$\pi\sqrt{\sigma/(\rho g)}$. For liquids we examine, namely, water,
ethanol and oil, gravitational effects are unimportant for scales
below $10\,$mm. The samples we examine have surface sizes of few mm
and longer wavelengths are effectively cut off, so that henceforth we
ignore effects due to gravity.
Under these circumstances, the dispersion relation reduces to the
following two equivalent relationships.
\begin{equation}
    \omega=\omr(k)=\sqrt{\sigma k^3\over \rho} 
    \qquad,\qquad
    k=\kr(\omega)= \left(\rho\omega^2\over\sigma\right)^{1/3} \qquad.
\end{equation}
In what follows, we examine the full spectrum both experimentally and
theoretically, including the full effects of dissipation.

The paper is organized as follows: In \sect{measurement}, we explain
the surface light reflection experiment and derive what precisely is
measured by this. The properties of thermal surface fluctuations of
simple liquids are examined in \sect{disp} and the approximate simple
analytic behavior of the spectra are derived. The shot noise level in
our experiments are assessed in \sect{shotNoise}. The noise reduction
method explained in \sect{noiseReduction} allows us to detect weak
signals buried under the shot noise level. The general applicability
of the principle is also clarified. Finally, we combine the
theoretical and the experimental results in \sect{exp} and find that
they agree. The limitations in the experiment and further directions
for research are also discussed.
\section{The experiment and the measurement}
\label{sec:measurement}

\begin{figure}[htbp]
    \centering
   \includegraphics[width=9cm,clip=true]{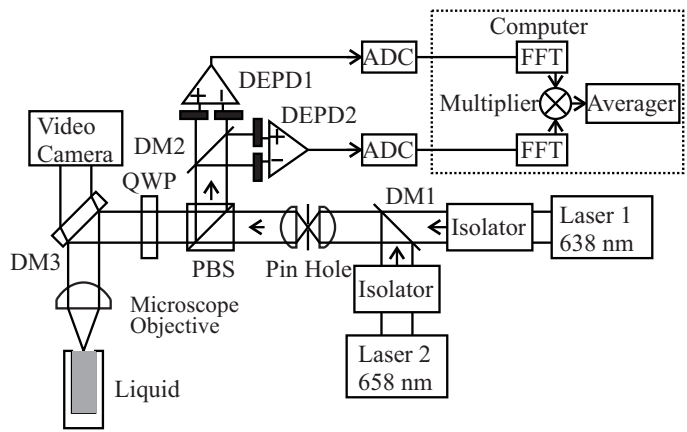}
   \caption{Two laser beams with wavelengths 638\,nm and 658\,nm are
     combined at a dichroic mirror (DM1) and focused by a microscope
     objective lens onto the liquid surface. The reflected light with
     different wavelengths are separated at DM2 to obtain two
     independent inclination measurements of the same surface. The
     inclination is converted into electric signals using DEPD1, DEPD2
     and are fed via ADC's into a computer where FFT (fast Fourier
     transform) and averagings are performed. A polarizing beam
     splitter (PBS) and a quarter wave plate (QWP) are included to
     extract the light reflected back from the sample efficiently.
     DM3 is used for viewing the sample through a video camera.  }
    \label{fig:setup}
\end{figure}

In the experiment (\figno{setup}), inclination fluctuations of a
liquid surface are measured using two sets of measurement systems
each with a laser beam 1 (wavelength 638 nm) and 2 (658 nm). The
average inclination within the beam area essentially acts as an
optical lever and the inclination deflects the laser beams to be
detected by the dual-element photodiodes (DEPD 1,2, S4204 Hamamatsu
Photonics, Japan)\cite{opticalLever}.  The two sets of photodiodes are
necessary here to eliminate the random noise statistically, as
explained in Sect.~\ref{sec:noiseReduction}.  The DEPD signals are
amplified and then fed into a computer via analog to digital
converters (ADC, 14 bit, ADXII14-80M, Saya, Japan).  Fourier
transforms and averagings are performed by the computer.

The laser beam power at the sample is 0.5\,mW each.  The beam is
focused down to $\mu $m order and its diameter ($\sim\lambda/\NA$) is
varied by changing the numerical aperture ($\NA$) of the objective
lens, where $\lambda$ is the wavelength of the probe laser beam. This
beam size is considerably smaller than that used in the standard light
scattering methods. The amplitude of the waves are small compared to
the wavelength so that the reflected light is almost all collected by
the objective lens. Therefore, compared to the standard light
scattering experiments which observes only a small fraction of the
scattered light, we obtain larger signals for a given beam power.
This can be crucial in low power measurements.

We now explain what is measured in this experiment and how this
relates to the spectrum of surface fluctuations.  We 
measure fluctuations in the average inclination of the
surface. Denoting the surface level as $\phi(\bm r,t)$, the average
inclination $a_1(t)$ can be obtained by approximating $\phi(\bm r,t)$
by a linear profile, minimizing the following quantity:
\begin{equation}
    \label{eq:lsq}
    \int d^2\bm r\, G(\bm r)\left|\phi(\bm 
r,t)-\left(a_0(t)+a_1(t)x\right)\right|^2
    \qquad.
\end{equation}
Here, $\bm r=(x,y)$ are coordinates on the surface and $G(\bm r)$ is
the beam intensity profile. This leads to the expression for the
average displacement and inclination,
\begin{eqnarray}
    \label{eq:aSol}
    a_0(t)&=&C_0\int d^2\bm r\,G(\bm r)\phi(\bm r,t)\quad,\quad
    C_0=\left( \int d^2\bm r\,G(\bm r)\right)^{-1} \qquad,\\
    a_1(t)&=&C_1\int d^2\bm r\,G(\bm r)x\phi(\bm r,t)\quad,\quad
    C_1=\left( \int d^2\bm r\,x^2G(\bm r)\right)^{-1} \qquad.
\end{eqnarray}
While we can solve for $a_j(t)$ more generally, we assumed here that
the profile is symmetric, $\int d^2\bm r\,xG(\bm r)=0$, which applies
to our case, as discussed below. We define the Fourier transform of
$a_j$ as
\begin{equation}
    \label{eq:FTa}
    \tilde a_j(\omega)={1\over\sqrt T}\int^{T/2}_{-T/2}\!\!dt\,
    e^{i\omega t}a_j(t)\quad,\qquad j=1,2 \qquad,
\end{equation}
where $T$ is the measurement time, which is much longer when compared
to the other time scales involved.

Since the correlation function of the surface fluctuations is
translation invariant both in space and time, it can be expressed as
\begin{equation}
    \label{eq:corr}
    \vev{\phi(\bm r,t)\phi(\bm r',t')}
    =\int{d\omega\over2\pi}\int{d^2\bm k\over(2\pi)^2}\,
    e^{i\bm k(\bm r-\bm r')-i\omega(t-t')}P(k,\omega)\qquad.
\end{equation}
Here, $P(k,\omega)$ is the spectral function of the fluctuations of
the surface displacement and $\vev{\cdots}$ denotes the statistical
average. Using this correlation function, the fluctuations in the
inclinations are obtained as
\begin{equation}
    \label{eq:awResult}
    \vev{|\tilde a_1 (\omega )|^2}
    =\int{d^2\bm k\over(2\pi)^2}\,
    \left|C_1\int d^2\bm r\,e^{i\bm k\bm r}xG(\bm
      r)\right|^2P(k,\omega)
    \qquad.
\end{equation}

Since we are using a laser beam well described by a Gaussian profile
to observe surface fluctuations, the profile function $G(\bm r)$ can
be expressed using  the beam diameter $\bsize$ as 
\begin{equation}
    \label{eq:GDef}
    G(\bm r)=I_0e^{-8r^2/\bsize^2}\qquad,
\end{equation}
The spectrum measured in the experiment is (we henceforth use the
notation $\omega=2\pi f$)
\begin{equation}
    \label{eq:spectrumF}
    S(f)=4\pi\vev{|\tilde a_1 (\omega )|^2}\qquad,
    % \quad, \qquad\omega=2\pi f
\end{equation}
taking into account that the measurement is in frequency space and the
one-sidedness of the spectrum. Combining the results above, we derive
a compact expression for the fluctuation spectrum observed in the
experiment,
\begin{equation}
    \label{eq:spectrum}
    S(f)=\int_0^\infty dk\,k^3e^{-\bsize^2k^2/16}P(k,2\pi f)\qquad.
\end{equation}
It should be noted that this formula applies to general surface
fluctuation spectra measured using this method and is not limited to
liquids nor to thermal fluctuations. This result specifies the
measured spectrum $S(f)$ {\it completely} including its magnitude,
given the spectral function $P(k,\omega)$ and the beam diameter
$\bsize$, and is independent of the beam power applied. As can be seen
from the expression, the role of the beam size is to effectively cut
off the $k$ integral of the spectral functions for values over
$\sim2\pi/\bsize$. This occurs because the inclination is effectively
averaged within the beam spot, so that shorter wavelengths are
effectively averaged out. It also explains why we integrate up to
infinity in this formula; while, in principle, the wavelengths of
surface fluctuations should be cutoff at atomic length scales, this is
much smaller than $\bsize$ so that using infinity as the upper limit
in the integration region introduces negligible difference, due to the
Gaussian damping.  The lower limit of the integration region should,
strictly speaking, be set to $\sim 2\pi/L$, where $L$ is the size of
the sample ($L$ is few mm in our experiment) providing an upper bound
for wavelengths. However, the difference from setting the lower end of
the integral to zero as in the formula can also be ignored, as will
become clear below.
\section{Analytic structure of the spectrum}
\label{sec:disp}
The spectral function for thermal fluctuations of simple liquid
surfaces has been derived previously\cite{Bouchiat},
\begin{equation}
    \label{eq:ripplon}
    P(k,\omega)={\kb T\over \pi}
    {ku^2\over \rho\omega^3}\Im\left[(1-iu)^2+y-\sqrt{1-2iu}\right]^{-1}
    ,\qquad
    u\equiv{\rho\omega\over 2\eta k^2}\quad
    y\equiv {\rho\sigma\over4\eta^2 k}\qquad.
\end{equation}
While this expression for the spectrum is analytic, its behavior is
not apparent. The behavior can be split into several regimes depending
on the viscosity. We summarize the behavior concisely below to gain
insight into the behavior of the spectral function and for later use.
\subsection{Leading analytic behavior with respect to $k$}
Let us study the behavior of the spectral function $P(k,\omega)$ with
respect to $k$ with $\omega$ fixed.
The dimensionless measure of viscosity of the liquid,
$\eta^3\omega/(\rho\sigma)$ influences the behavior of the spectral
function $P(k,\omega)$ qualitatively.  Any liquid is highly dissipative
for high enough frequencies, which is intuitively natural.  When the
viscosity is effectively low, $\eta^3\omega/(\rho\sigma)<1/(8\sqrt2)$,
the leading order analytic behavior $P_0(k,\omega)$ 
can be obtained as
\begin{equation}
    \label{eq:p01}
    P_0(k,\omega)={\kb T\over\pi}\times\left\{
    \begin{array}[c]{lll}
        \displaystyle{{4\eta k^3\over\rho^2\omega^4}}&
        \qquad\hbox{when }&k<2^{-1/6}\kr(\omega)\\
        &&\\
        \displaystyle{{2\eta\over\sigma^2k^3}}
        && k>2^{-1/6}\kr(\omega)\\
    \end{array}\right.
  \qquad.
\end{equation}
In this case, $P(k,\omega)$ has a peak close to $k=\kr(\omega)$, which
becomes less prominent as $\eta^3\omega/(\rho\sigma)$ increases.
When $\eta^3\omega/(\rho\sigma^2)>1/(8\sqrt2)$, the leading behavior of
the dispersion relation splits into three regions as
\begin{equation}
    \label{eq:p02}
    P_0(k,\omega)={\kb T\over\pi}\times\left\{
    \begin{array}[c]{lll}
        \displaystyle{{4\eta k^3\over\rho^2\omega^4}}
         &\qquad\hbox{when }&
         \displaystyle{k<2^{-1/4}\sqrt{\rho\omega\over2\eta}}\\
        &&\\
        \displaystyle{{1\over2\eta k\omega^2}}&
        &\displaystyle{
           2^{-1/4}\sqrt{\rho\omega\over2\eta}<k<{2\eta\omega\over\sigma}}\\
        &&\\
        \displaystyle{{2\eta\over\sigma^2k^3}}&
         & \displaystyle{k>{2\eta\omega\over\sigma}}\\
    \end{array}\right.
  \qquad.
\end{equation}
In both cases with low and high viscosity, for long wavelengths
compared to $2\pi/\kr(\omega)$, the spectral function is governed by
viscous behavior and depends on $\rho$. On the other hand, for short
wavelengths, it is suppressed by the surface tension of the liquid and
is independent of its density.

\begin{figure}[htbp]
    \centering
   \includegraphics[width=\plotX,clip=true]{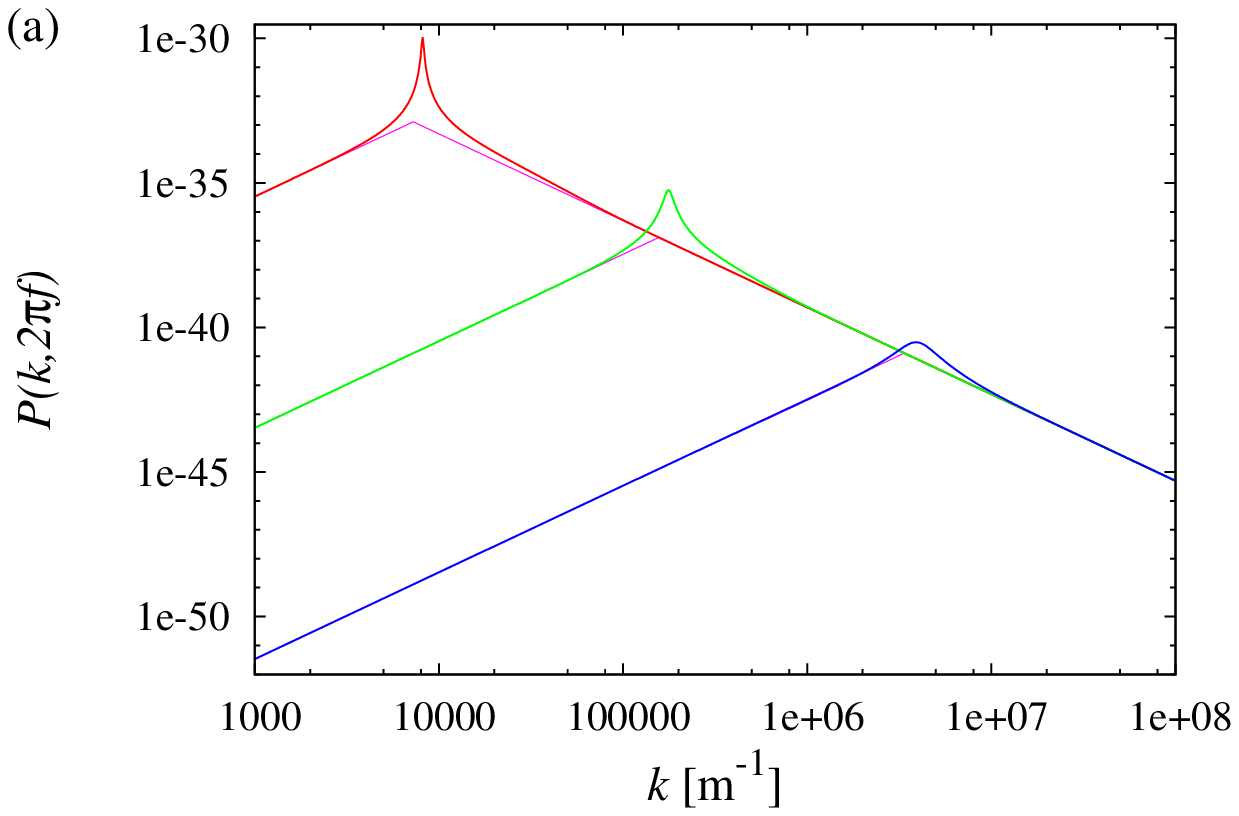}
   \includegraphics[width=\plotX,clip=true]{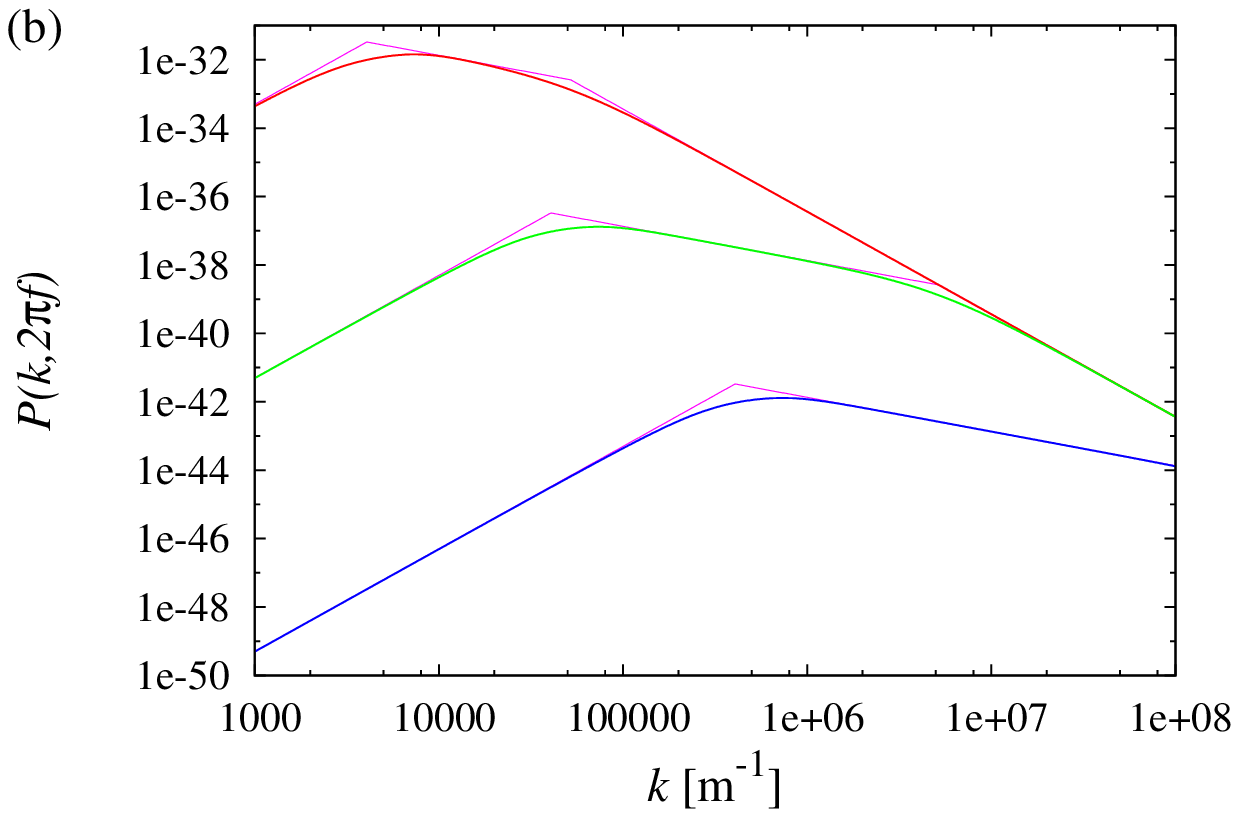}
   \caption{(Color online) Behavior of $P(k,2\pi f)$ for (a) water and
     (b) oil with respect to $k$ at fixed $f$. Frequencies are $
     f=10^3, 10^5, 10^7$\,[s$^{-1}$] (red, green, blue lines
     respectively), with the maximum of the spectrum being larger for
     higher $f$.  The corresponding simple analytic behaviors
     Eqs.~\eqnn{p01},\eqnn{p02} are also shown (thin lines, magenta)
     and matches well with the full spectral function and are almost
     invisible, except at the boundaries between the regions. }
    \label{fig:dispK}
\end{figure}

In \figno{dispK}, we compare the full spectral function
Eq.~\eqnn{ripplon} with its approximate analytic behavior derived
above. We see that in all the cases, the spectrum is well reproduced
by the simple analytic behaviors, except for the peak seen in the water
surface fluctuations at lower frequencies. The peak behavior reflects
long lived waves and disappears when the viscosity is effectively
high, due to dissipation.  In both Eqs.~\eqnn{p01}, \eqnn{p02}, the
spectral function is independent of $f$ for large $k$, which can be
seen from the plots. The simple analytic formulas derived above capture
the situations with weak and strong viscosity, which have
qualitatively different behaviors.
\subsection{Leading analytic behavior with respect to $\omega$}
\label{sec:analy}
We now analyze the behavior of $P(k,\omega)$ with respect to $\omega$
for fixed $k$.  For a liquid with low viscosity effectively,
$\eta^2k/(\rho\sigma)<1/(4\sqrt2)$, the leading order analytic
behavior $P_0(k,\omega)$ is broken up into two regions as
\begin{equation}
    \label{eq:pk01}
    P_0(k,\omega)={\kb T\over\pi}\times\left\{
    \begin{array}[c]{lll}
        \displaystyle{{2\eta\over\sigma^2k^3}}
        &\qquad\hbox{when }&\omega<2^{1/4}\omr(k)\\
        &&\\
        \displaystyle{{4\eta k^3\over\rho^2\omega^4}}
        &&\omega>2^{1/4}\omr(k)\\
    \end{array}\right.
  \qquad.
\end{equation}
For the highly viscous case, the leading analytic behavior can be
broken down into three regions thus.
\begin{equation}
    \label{eq:pk02}
    P_0(k,\omega)={\kb T\over\pi}\times\left\{
    \begin{array}[c]{lll}
        \displaystyle{{2\eta\over\sigma^2k^3}}
        &\qquad\hbox{when }&\displaystyle{\omega<{\sigma k\over2\eta}}\\
        &&\\
        \displaystyle{{1\over2\eta k\omega^2}}
        &&\displaystyle{{\sigma k\over2\eta}<\omega
          <{2\sqrt2\eta k^2\over \rho}}\\
        &&\\
        \displaystyle{{4\eta k^3\over\rho^2\omega^4}}
        &&\displaystyle{\omega>{2\sqrt2\eta k^2\over \rho}}\\
    \end{array}\right.\qquad.
\end{equation}
\begin{figure}[htbp]
    \centering
   \includegraphics[width=\plotX,clip=true]{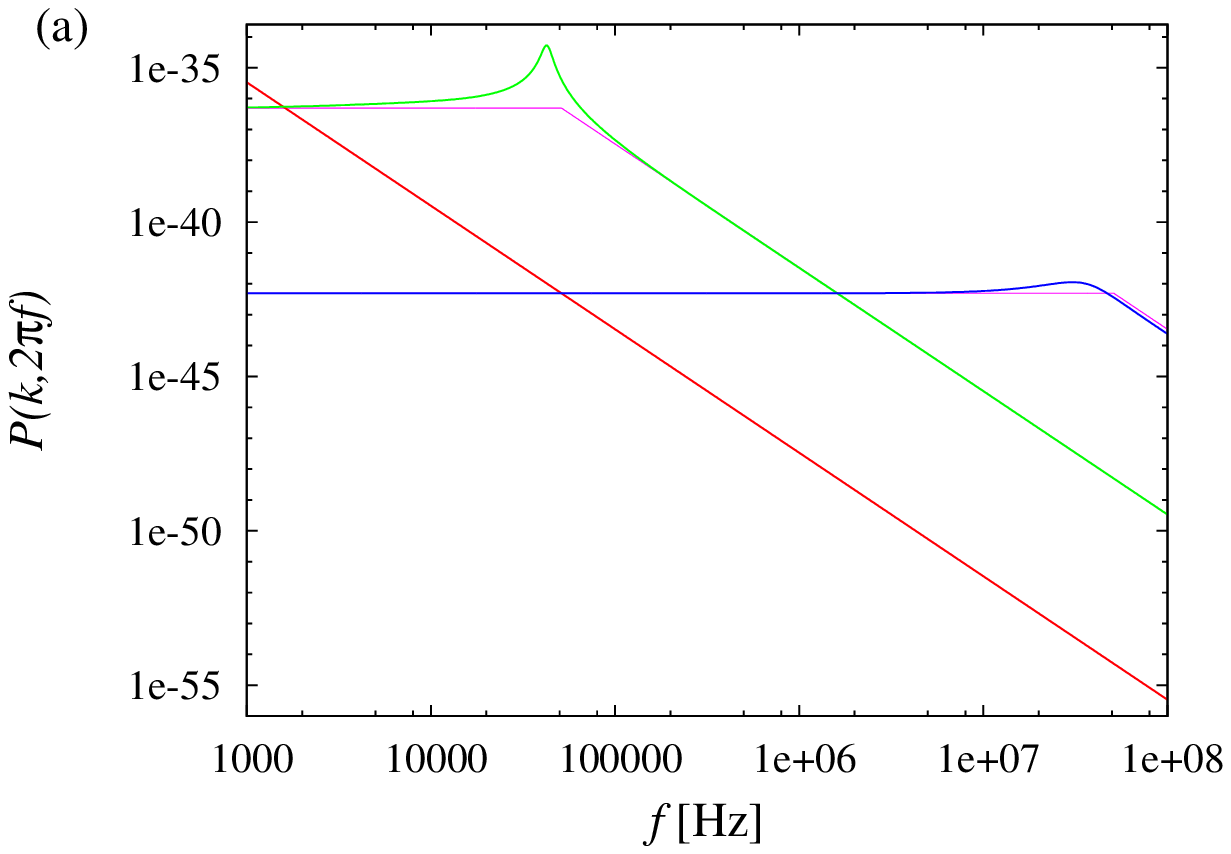}
   \includegraphics[width=\plotX,clip=true]{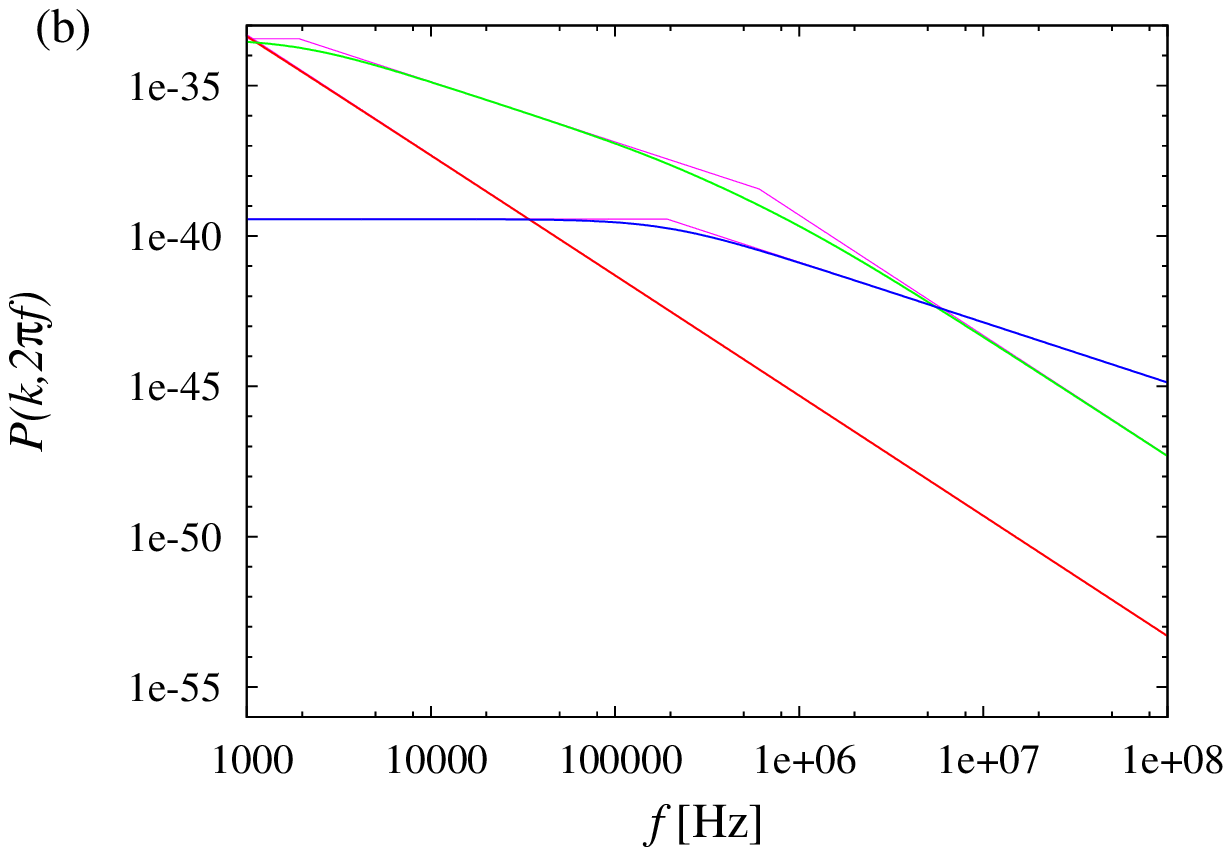}
   \caption{(Color online) Behavior of $P(k,2\pi f)$ for (a) water and
     (b) oil with respect to frequency $f$ at
     $k=10^3,10^5,10^7$\,[m$^{-1}$] (red, green, blue lines
     respectively). The spectra with larger $k$ have more fluctuations
     at higher $f$. Their analytic approximations Eqs.~\eqnn{pk01},
     \eqnn{pk02} (thin lines, magenta) agree well with the full
     spectral function and are almost invisible, except at the
     boundaries between the regions. }
    \label{fig:dispW}
\end{figure}
\subsection{Analytic behavior of the integrated spectrum}
\label{sec:spec}
While the spectrum $S(f)$ in Eq.~\eqnn{spectrum} can be computed
numerically, we now clarify its rough analytic behavior using
$P_0(k,2\pi f)$ derived above.  In the integrated spectrum $S(f)$, $b$
provides a cutoff $\kMax$ for the integral, whose relation we specify
below. The spectrum can be approximated by
\begin{equation}
    \label{eq:spectrumA}
    S_0(f)=\int_0^{\kMax} dk\,k^3 P_0(k,2\pi f)\qquad.
\end{equation}
This can be computed explicitly. For the low viscosity case
$\eta^3\omega/(\rho\sigma^2)<1/(8\sqrt2)$,
\begin{equation}
    \label{eq:SALow}
    S_0(f)={\kb T\over\pi}\times\left\{
    \begin{array}[c]{lll}
          \displaystyle{\left({2\eta\kMax\over\sigma^2}-{2^{11/6}3
              \rho^{1/3}\eta\over7\sigma^{7/3}}\omega^{2/3}\right)}
          &\quad\hbox{when }&\displaystyle{\omega<2^{1/4}\omr(\kMax)}\\
        &&\\
        \displaystyle{{4\eta\kMax^7\over7\rho^2}{1\over\omega^4}}
        &&\displaystyle{\omega>2^{1/4}\omr(\kMax)}\\
    \end{array}\right.\qquad.
\end{equation}
When the viscosity is high, the spectrum has the following
approximate.
\begin{equation}
    \label{eq:SAHigh}
    S_0(f)={\kb T\over\pi}\times\left\{
    \begin{array}[c]{lll}
          \displaystyle{\left({2\eta\kMax\over\sigma^2}
              -{8\eta^2\over3\sigma^3}\omega
            -{\rho^{3/2}\over2^{5/4}21\eta^{5/2}}{1\over\omega^{1/2}}\right)}
          &\quad\hbox{when }&\displaystyle{\omega<{\sigma\kMax\over2\eta}
            }\\
            &&\\
          \displaystyle{\left({\kMax^3\over6\eta}{1\over\omega^2}
            -{\rho^{3/2}\over2^{5/4}21\eta^{5/2}}{1\over\omega^{1/2}}\right)}
          &&
          \displaystyle{{\sigma\kMax\over2\eta}
            <\omega<{2\sqrt2\eta\kMax^2\over\rho}}\\
          &&\\
        \displaystyle{{4\eta\kMax^7\over7\rho^2}{1\over\omega^4}}
          &&\displaystyle{\omega>{2\sqrt2\eta\kMax^2\over\rho}
            }\\
    \end{array}\right.\qquad.
\end{equation}
Taking into account the formula for $S(f)$ in Eq.~\eqnn{spectrum}, we use
$\kMax=2^{1/4}4/b$, which satisfies $\int_0^\infty
dk\,\exp(-b^2k^2/16)k^3=\int_0^\kMax dk\,k^3$.
In \figno{dispW}, the spectrum $S(f)$ is compared to its approximate
analytic behavior $S_0(f)$ in Eqs.~\eqnn{SALow}, \eqnn{SAHigh}.
It can be seen that the essential features of $S(f)$ are well
reproduced by the simple analytic formulas. Examining in more detail,
we see that the agreement is better for more viscous liquids. This is
presumably due to the peak contribution unaccounted for in the
formulas Eqs.~\eqnn{p01}, \eqnn{p02}, which do not exist for more viscous
fluids. It can be seen from Eqs.~\eqnn{SALow}, \eqnn{SAHigh} that $S(f)$
behaves as $\sim\eta/\sigma^2$ for lower frequencies explaining why
oil has larger fluctuations than water in this regime. Also, from the
formulas, we can see why the fluctuations are larger for smaller $b$
(or equivalently, larger $\kMax$) over the whole spectrum, both when
the viscosity is weak and strong.

We note that the calculations above also show why the integration
region in the spectrum formula Eq.~\eqnn{spectrum} can be taken down to
zero as long as the cutoff is well below the upper limit of the
region, since had we put in a lower cutoff $k_{\rm low}$, its
contribution to $S(f)$ would behave as $\sim k_{\rm low}^7$. This
condition is equivalent to the surface size being much larger than the
beam diameter, which is always satisfied in our experiments.
\begin{figure}[htbp]
    \centering
   \includegraphics[width=\plotX,clip=true]{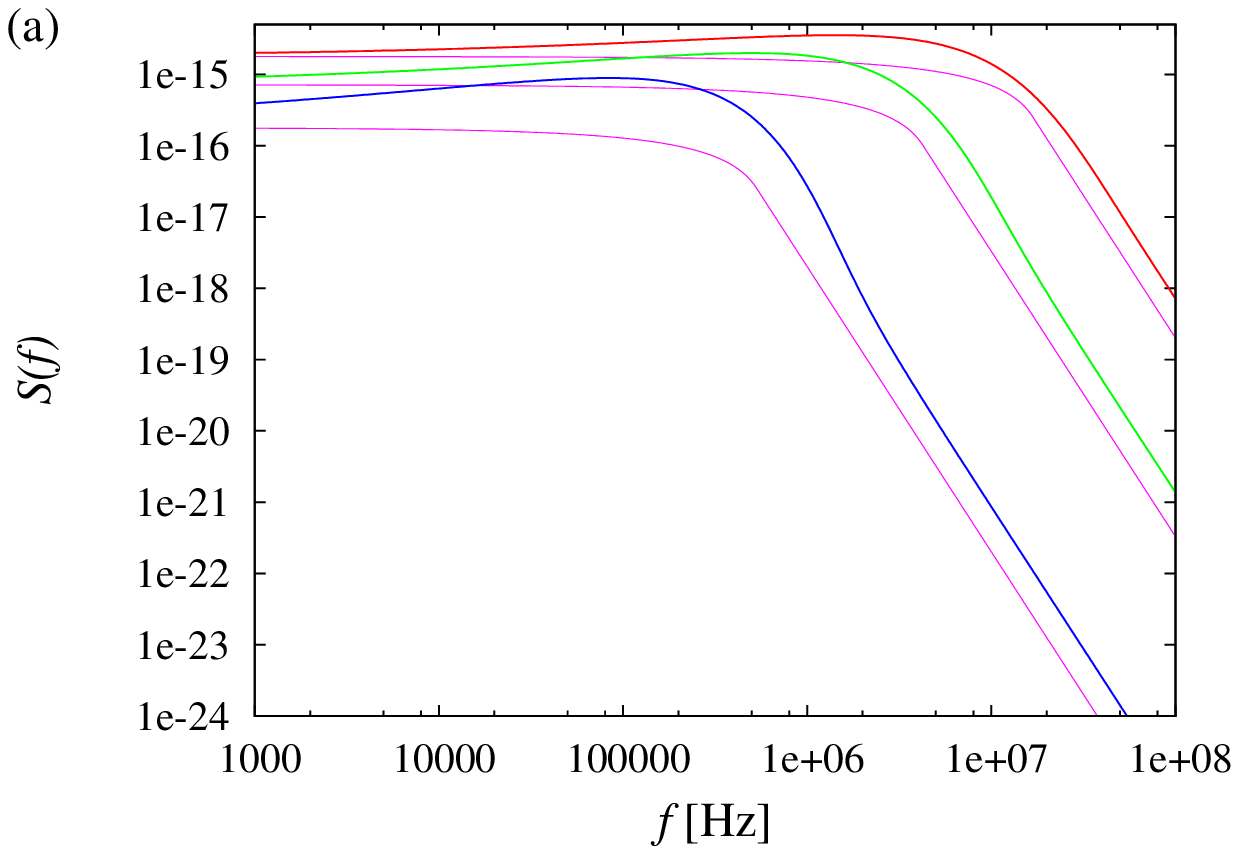}
   \includegraphics[width=\plotX,clip=true]{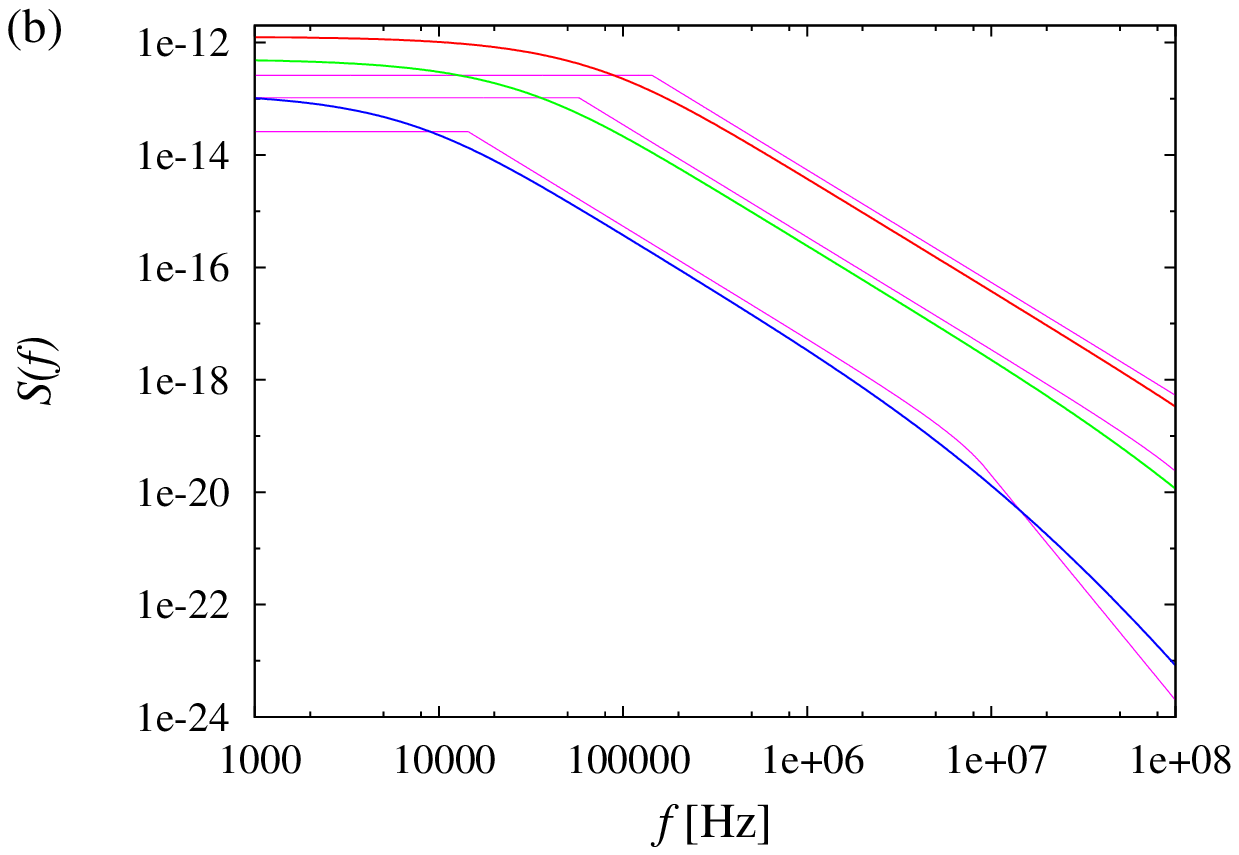}
   \caption{(Color online) Behavior of the integrated spectra of
     surface fluctuations $S(f)$ for (a) water and (b) oil with
     respect to $f$. The spectra are shown for
     $b=1,2.5,10\,[\mu$m$^{-1}$] (red, green, blue lines
     respectively), along with their analytic approximations (thin
     lines, magenta). The spectra for smaller $b$ have larger
     fluctuations across the spectrum.}
    \label{fig:dispW}
\end{figure}
\section{Shot noise level}
\label{sec:shotNoise}
While our measurement precision is not limited by the shot noise of
the probe laser beam, it is of import to understand how the shot noise
level would appear in our setting.  The electric current through the
photodiode is $ I=\beff \elec \bpower/ h\nu$,
where $\bpower$ is the signal power, $\elec $ is the electron charge
magnitude and $\beff$ is the quantum efficiency of the photodiode.
$\beff\simeq0.8$ for the photodiodes we use. The current generated by
shot noise is
\begin{equation}
    \label{eq:sn}
    I_{\rm SN}=\sqrt{2\elec I\,\Delta f}
    =\elec \sqrt{{2\beff \bpower\over h\nu}\Delta f}\qquad.
\end{equation}
$\Delta f$ is the frequency range of the measurement.  The signal in
the experiment comes from difference in the current $\Delta I$ due to
the geometric effects of the optical lever. For an inclination
$\theta$, the signal is
\begin{equation}
    \label{eq:signal}
    \Delta I
    % =\beff {\elec\Delta \bpower \over h\nu}
    =\beff{2\elec\theta \bpower\over    h\nu \NA}\qquad.
\end{equation}

The shot noise level in our experiment is the size of the angular
fluctuations $\theta^2_{\rm SN}$ corresponding to the shot noise
current. This is what appears in the measurements, had we {\it not}
used the noise reduction through correlations described in
\sect{noiseReduction}.  $\theta^2_{\rm SN}$ can be obtained from
$\Delta I\simeq I_{\rm SN}$ to be
\begin{equation}
    \theta_{\rm SN}^2 \simeq {\NA^2 \elec \over 2I}\Delta f\qquad.
\end{equation}
Here, the current $I$ is the photoelectric current collected by the
dual photodiodes.  $\theta_{\rm SN}^2$ is smaller for larger signals,
which is natural. Also, for larger numerical apertures, the beam spot
is focused down to a smaller size so that the effect of the optical
lever is smaller, hence the shot noise effects are larger.  In our
experiment, $I\simeq0.5\,\mu$A so that $\theta^2_{\rm SN}\simeq
\NA^2\times1.6\times10^{-13}\rm\,Hz^{-1}$.
It is crucial to separate out the signal from this noise, using
methods explained in the next section.
\section{Noise reduction through correlations}
\label{sec:noiseReduction}
An essential feature of thermal fluctuations is that they are random.
Since our objective is to directly measure the fluctuations under
normal circumstances, the fluctuations are furthermore small.  Even in
ideally executed experiments, some random noise, such as shot noise,
always exist. Therefore, to measure weak random signals, we need to
separate out the random signal from the random noise. This is possible
under rather general circumstances, as we now explain\cite{am1}.  The
principle behind this noise reduction is not limited to thermal
fluctuations or optical measurements, but applies generally to the
extraction of random signals from random noise. The conditions for its
applicability will be discussed below.

A detector measurement $D_1=S+N_1$ consists of the desired signal $S$
and some noise $N_1$, independent of $S$. Denoting Fourier transforms
with tildes, the power spectrum obtained under simple averaging is
\begin{equation}
    \label{eq:simpleAverage}
    \vev{|\tilde D_1|^2} = \vev{|\tilde S|^2} + \vev{|\tilde
      N_1|^2}  \qquad.  
\end{equation}
Since the signal itself is random in nature, there is no way to
distinguish the signal from the noise, so that the signal can not be
measured, unless the signal is larger than the noise, $\vev{|\tilde
  S|^2} \gg \vev{|\tilde N_1|^2}$. 
If we use only one measurement, this is an essential limitation. 

To overcome this obstacle, we make another independent measurement of
the same signal, $D_2=S+N_2$, where $N_2$ denotes the noise for this
measurement.  Then,
\begin{equation}
    \label{eq:decorr}
    \vev{\overline{\tilde D_1}\tilde D_2}\rightarrow \vev{|\tilde S|^2} 
\quad,
    \qquad{\ns}\rightarrow\infty    \qquad,
\end{equation}
eliminating the random noise $N_1,N_2$. Here $\cal N$ is the number of
averagings. This results holds since the measurements $D_1,D_2$ are
independent and the cross-terms of decorrelated random observables
(and their Fourier transforms) vanish under averaging.  The relative
error in this method is $\sim1/\sqrt{\ns}$, which arises from the
statistical nature of the method. In principle, given enough
averagings, we can suppress shot noise and other random noise effects
to an arbitrarily small size.

The crucial requirements for our method to work is that multiple
independent measurements can be made and that the signal is stable
enough to withstand averagings. The independence of the measurements,
or equivalently, the decorrelation of the noise in them is clearly
crucial. As the signal becomes weaker, stricter independence is
required, which in practice can be quite delicate. For instance, cross
talks can arise in electronic circuits, unless they are completely
separated and electronic signals can affect each other through
electromagnetic fields in the intervening space. These properties put
practical limitations on the reduced noise level. While our method can
not be used for an one-time event, it can be used for any recurrent
signal.

Another possible approach to reducing the relative noise is to
increase the signal strength. In our context, this would mean
increasing the beam intensity. However, this is not always applicable,
since a stronger beam will affect the sample. Even for simple liquid
surfaces, it leads to more evaporation and can lead to less precise
measurements. More generally, if we consider biological materials or
medical applications\cite{Denk}, using a strong light source is
often excluded.
Correlation measurements have been used previously in surface light
scattering experiments\cite{Earnshaw}. Our approach differs from those
in that we use the cross-correlation of independent measurements of
the same signal to reduce the noise.
\section{Experimental results and theory}
\label{sec:exp}
\begin{figure}[htbp]
\noindent\includegraphics[width=\plotX,clip=true]{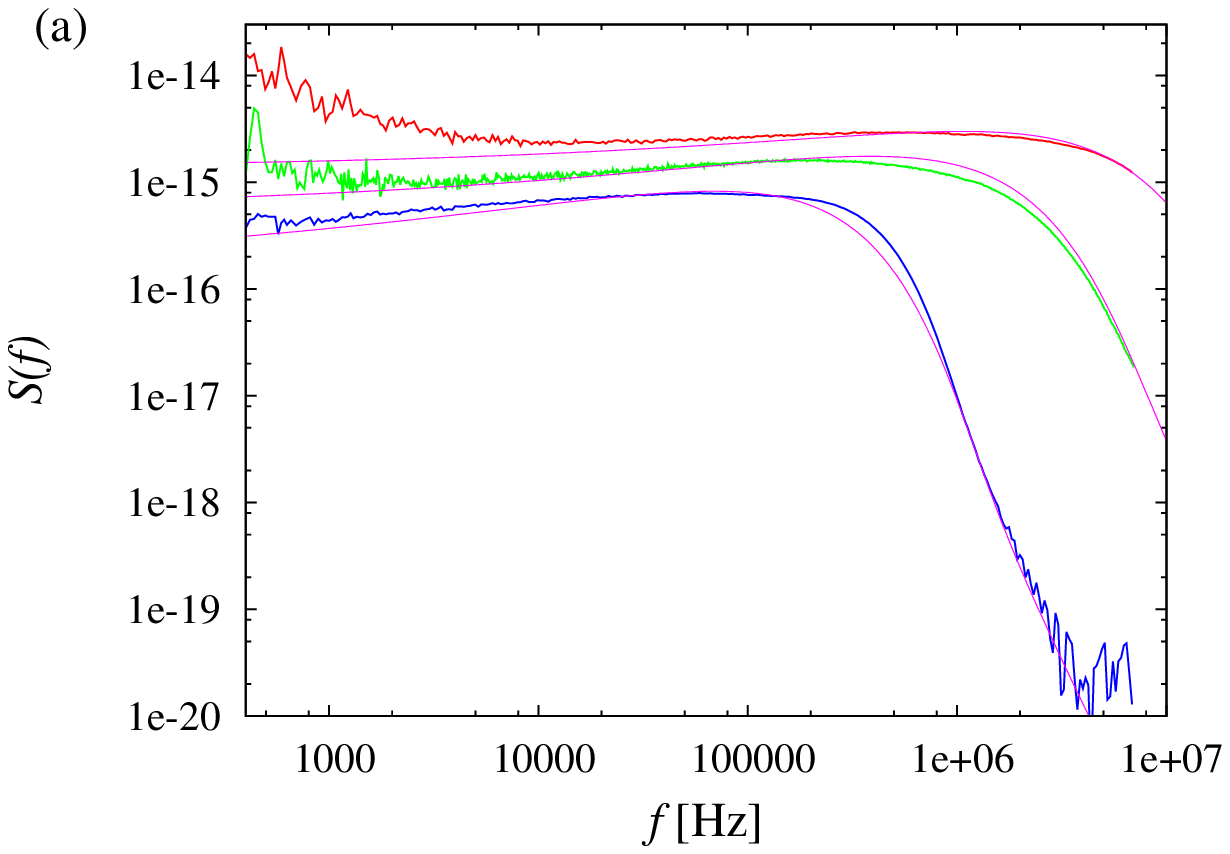}
    \includegraphics[width=\plotX,clip=true]{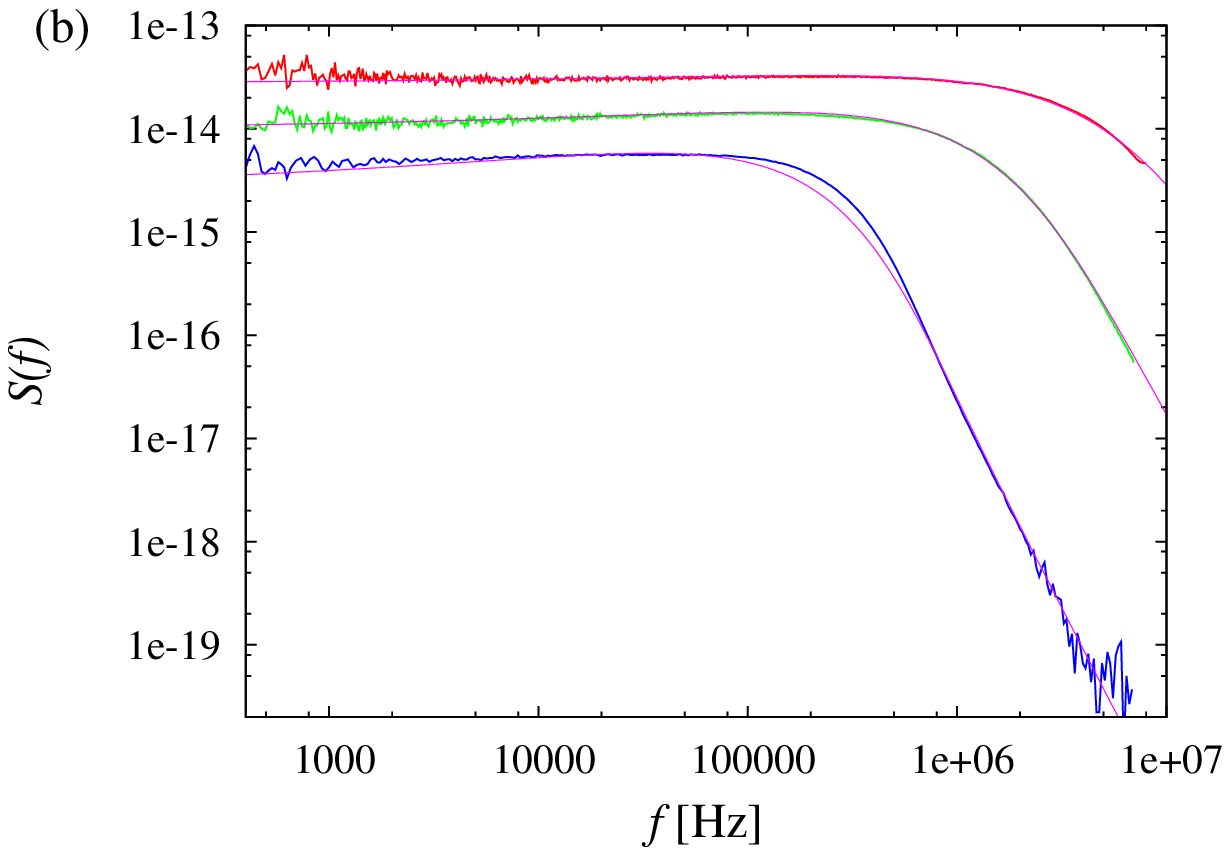}\\
\noindent\includegraphics[width=\plotX,clip=true]{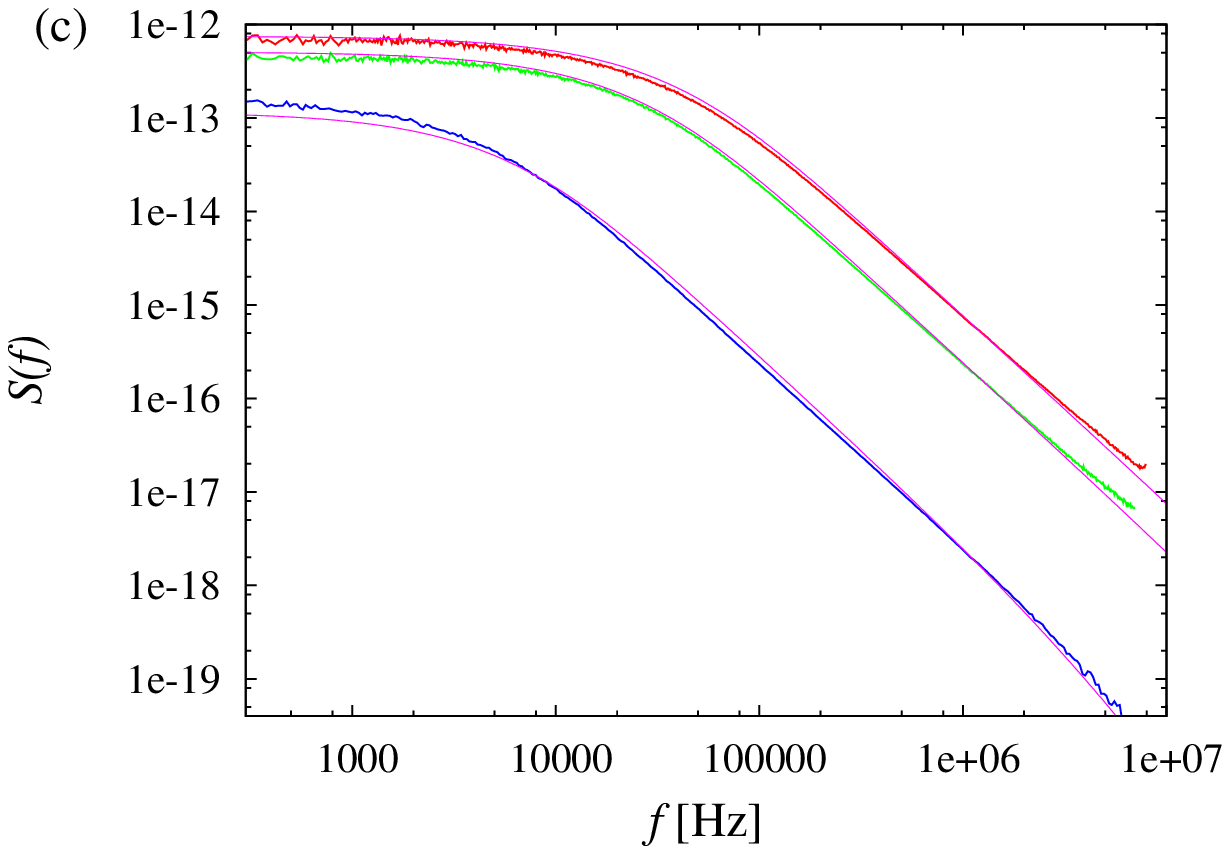}
\caption{(Color online) Observed fluctuation spectra for (a) water
  ($b=1.3, 3.1, 11.5\,[\mu\rm m]$), (b) ethanol ($b=0.9, 2.5,
  9.5\,[\mu\rm m]$) and (c) oil ($b=1.7, 2.5, 11.0\,[\mu\rm m]$),
  (red, green, blue lines from smaller to larger $b$). Respective
  theoretical spectra $S(f)$ are also shown (thin lines, magenta).
  The spectra for smaller $b$ have larger fluctuations.  }
    \label{fig:exp}
\end{figure}

In \figno{exp}, we compare the experimental results with the theory
explained above for water, ethanol and oil with various beam sizes.
The fluid properties we used for water, ethanol and oil are
$(\rho\,{\rm [kg/m^3]}, \sigma\,{\rm [kg/s^{2}]},\eta\,{\rm [kg/(
  m\cdot s)]})=(1.0\times10^3,7.3\times10^{-2},1.0\times10^{-3})$,
$(0.79\times10^3,2.2\times10^{-2},1.1\times10^{-3})$ and
$(0.92\times10^3,3.0\times10^{-2},0.124)$, respectively.  The
agreement between the theoretical formula Eq.~\eqnn{spectrum} and the
experimental measurements is mostly quite satisfactory, including its
beam size dependence. There is some excess signal in the water surface
fluctuations at low frequencies for small $b$ (b=1.3\,$\mu$m case in
\figno{exp}(a)), whose cause is explained below. The rough features of
the spectra can be understood from Eqs.~\eqnn{SALow} and
\eqnn{SAHigh}; the spectrum behaves as $S(f)\sim\eta/(\sigma^2\bsize)$
at low frequencies. So, the fluctuations are the largest for oil due
to its large viscosity and smallest for water due to its large surface
tension. The fluctuations are larger for smaller $\bsize$.  At high
frequencies, as we decrease $\bsize$, the effective cutoff for the
wavelength decreases and fluctuations at larger frequencies become
more apparent.

The excess measurements in the water surface fluctuation spectrum at
low frequencies for $b=1.3\,\mu$m can be explained as follows:
Ideally, the experiment measures only inclination fluctuations of the
surface. However, given the high sensitivity of the measurements,
vertical displacements can mimic inclinations due to the focusing of
the objective lens, when the beam alignment is not perfect.  This
effect is contained in $\vev{|\tilde a_0(\omega)|^2}$ which behaves as
$ \sim1/f$ at low frequencies.  For smaller $b$, this effect is larger
due to the steeper focusing and can be seen in the water surface
fluctuation measurements at low frequencies, since for a given beam
size, $S(f)$ is smallest for water.

The shot noise levels in the measurements are $6\times10^{-14}, \
3\times10^{-14}, 2\times10^{-15}\,[\rm Hz^{-1}]$ for the beam
diameters $b=1.5, 2.5, 10\,[\mu$m], as explained in
\sect{shotNoise}. Therefore, we see that the noise reduction using
signal correlations explained in the previous section is crucial for
examining even the qualitative features of the spectra, since most,
and in some cases all, of the spectra is below shot noise levels.

Some comments regarding the calibration of the measurements is in
order.  On the theoretical side, given the properties of the liquid
and the beam diameter, there are {\it no} further parameters at all in
the spectrum Eq.~\eqnn{spectrum} and is specified completely.
Experimentally, the frequency dependence of the spectrum can be
measured precisely. Calibrating the overall magnitude of the measured
spectrum is more difficult and this is done using a piezoelectrically
driven mirror with a known oscillation amplitude. While this works
well when $b$ is large, for smaller $b$, the shallowness of the depth
of field makes the calibration less accurate.  Another complication is
that the liquid in general evaporates during the measurement so that
the beam defocuses. This, in effect, increases the beam diameter and
can influence the spectral shape. This problem is clearly more acute
for larger beam powers.
A typical measurement of simple liquid surface fluctuations takes
around 20 seconds and the beam power applied is $0.5\,$mW.
Given the excellent agreement of the observed fluctuation spectra of
simple liquid surfaces and theory, it might be reasonable to use
thermal surface fluctuations of a well studied specific liquid such as
oil to finely calibrate the system when applying the measurements to
more general samples.

In this work, we directly studied the spectra of thermal fluctuations
for simple liquids, using surface light reflection methods. The
spectra obtained are integrated over wavelengths and we have also
investigated the dependence of the spectra on the beam diameter. In
the process, we applied a novel general method for noise reduction,
using the correlation of independent measurements of the same
signal. The spectra obtained matches well with theory, whose analytic
behavior can be summarized rather simply. The measurement method is
complementary to the surface fluctuation measurements for specific
wavelengths, when applied to simple liquids. The method, especially
when combined with the noise reduction method, has a broad range of
applicability since the required sample size, observation time and
power are small.

\end{document}